\documentclass[aps,prd,onecolumn,groupedaddress,preprintnumbers,superscriptaddress,noshowpacs,nofootinbib]{revtex4}
\usepackage{amsmath,amssymb,latexsym}
\usepackage[dvipdfmx]{graphicx}
\usepackage{color}
\usepackage{mathrsfs}

\usepackage{enumerate}

\begin{document}

\title{
Extension of Kodama vector and quasilocal quantities in three-dimensional axisymmetric spacetimes
}

\author{Shunichiro Kinoshita}
\email{kinoshita@phys.chuo-u.ac.jp}
\affiliation{
Department of Physics, Chuo University, 
Kasuga, Bunkyo-ku, Tokyo 112-8551, Japan
}


\date{\today}

\begin{abstract}
 Spherically symmetric spacetimes admit the so-called Kodama vector,
 which provides 
 a locally conserved current and a preferred time even for dynamical
 spacetime without any time translation symmetry.
 A charge associated with this conserved current leads to a quasilocal
 mass which agrees with the Misner-Sharp mass.
 In three dimensions, spherically symmetric spacetimes correspond to
 axisymmetric ones, while axisymmetry allows
 spacetimes to be rotating with angular momentum.
 We extend the notion of the Kodama vector to axisymmetric rotating
 spacetimes in three dimensions.
 We also define a quasilocal mass taking into account angular momentum in
 three-dimensional axisymmetric spacetimes. 
\end{abstract}

\maketitle

 \section{Introduction}

 Conservation laws of energy and momentum have played significant roles in physics.
 In gravitational theories, however, diffeomorphism invariance makes 
 local notion of energy and momentum for gravitating system ambiguous.
 Thus, gravitational energy-momentum should be defined nonlocally in finite
 spacetime domain.
 Such various definitions of quasilocal mass and angular momentum have been
 proposed (see \cite{Szabados:2009eka,Jaramillo:2010ay} and references
 therein). 

 On the other hand, it is well-known that if a spacetime admits a Killing
 vector field generating some symmetry, a locally conserved
 current vector can be obtained by contracting divergence-free,
 symmetric tensor such as energy-momentum tensor, or Einstein tensor, with
 the Killing vector.
 In particular, if a Killing vector is timelike, 
 one can obtain a locally conserved energy current in terms of the
 Killing vector.
 In this case the spacetime is static or stationary with time
 translation symmetry, so that this conserved current cannot be applied
 to dynamics of gravitating systems.

 Remarkably, in spherically symmetric spacetimes we can construct a
 local conserved energy current by using the so-called Kodama vector,
 even though the spacetime is dynamical without any time translation
 Killing vector~\cite{Kodama:1979vn}.
 This conserved current gives us quasi-local mass as Misner-Sharp mass~\cite{Misner:1964je,Hayward:1994bu}.
 Generalizations to higher-dimensional spacetime or other gravitational
 theories such as Gauss-Bonnet and $f(R)$ gravity were studied in Refs.~\cite{Maeda:2006pm,Maeda:2007uu,Cai:2009qf,Zhang:2014goa}.

 In three dimensions, the counterparts of spherically symmetric spacetimes
 are axisymmetric (or circularly symmetric) ones.
 However, since axisymmetry does not rule out rotating systems, which are
 of physical interest significantly, we can take account of
 angular momentum unlike spherical symmetry of higher-dimensional spacetimes.
 Such a rotating spacetime, in general, cannot be described by
 warped product metric.
 It turns out that in spacetimes with angular momentum the conventional Kodama
 vector does not always yield a conserved current.
 (The Kodama vector in warped product spacetimes was discussed in Ref.~\cite{An:2017wti}.)

 In this paper we will extend the notion of the Kodama vector to
 spacetimes with nonzero angular momentum, which are not described by
 warped product metric.
 The organization of the paper is as follows.
 In Sec.~\ref{sec:extension} we consider general axisymmetric spacetimes
 in three dimensions, and explore a new vector that satisfies similar
 conditions to the Kodama vector in spherically symmetric cases.
 This new vector provides a locally conserved current.
 In Sec.~\ref{sec:quasi-local_quantities} we define quasilocal
 quantities as charges associated with locally conserved currents.
 
 While this paper was in preparation, Ref.~\cite{Gundlach:2021six}
 appeared where the authors have proposed the same quasilocal mass.

 \section{Extension of Kodama vector}
 \label{sec:extension}
 
 Let us consider a three-dimensional axisymmetric spacetime with 
 an axial Killing vector field given by 
 \begin{equation}
  \psi^\mu = \left(\frac{\partial}{\partial \phi}\right)^\mu .
 \end{equation}
 The spacetime metric can be generally written as 
 \begin{equation}
  g_{\mu\nu}dx^\mu dx^\nu =
   h_{ij}(y)dy^i dy^j + r^2(y) [d\phi + a_i(y)dy^i]^2 ,
 \end{equation}
 where $h_{ij}$ is a two-dimensional metric on the orbit space of the
 Killing vector.
 Note that 
 $a_i(y)$ has a gauge freedom as $a_i(y) \to a_i(y) + \partial_i \lambda(y)$,
 where $\lambda(y)$ is an arbitrary function on the orbit space.
 If we assume that the axial Killing vector field has closed orbits with
 $\phi \sim \phi + 2\pi$, 
 the norm of the Killing vector provides 
 the circumferential radius $r(y) = (\psi^\mu \psi_\mu)^{1/2}$.
 Greek indices indicate components of total three-dimensional spacetime, 
 and Latin indices indicate two-dimensional orbit space components.

 When $\psi^\mu$ is a hypersurface orthogonal Killing vector
 $\psi_{[\mu}\nabla_\nu \psi_{\lambda]} = 0$, that is, 
 nontrivial $a_i(y)$ vanishes,
 the spacetime becomes simply a warped product.
 In this case we obtain the Kodama vector by conventional
 definition~\cite{Kodama:1979vn} as $\widetilde{K}^i = - e^{ij}\nabla_j r$, 
 where $e^{ij}$ is the two-dimensional volume form associated with $h_{ij}$.
 This definition can be naturally adapted to a vector on the full
 three-dimensional spacetime as 
 \begin{equation}
  \widetilde{K}^\mu = - \frac{1}{r}\epsilon^{\mu\nu\alpha}\psi_\alpha
   \nabla_\nu r , 
 \end{equation}
 where $\epsilon_{\mu\nu\alpha}$ is the totally antisymmetric $3$-tensor
 corresponding to a three-dimensional volume form.
 The metric of the two-dimensional orbit space is written as 
 \begin{equation}
  h_{\mu\nu} = g_{\mu\nu} - \frac{1}{r^2}\psi_\mu \psi_\nu .
 \end{equation}

 Even though the spacetime is not a warped product, 
 $\widetilde{K}^\mu$ can still satisfy the following properties:
 $\widetilde{K}^\mu \nabla_\mu r = 0$ and 
 $\nabla_\mu \widetilde{K}^\mu = 0$ in a similar manner as the Kodama vector.
 However, it turns out that 
 $G^{\mu\nu}\nabla_\mu \widetilde{K}_\nu \neq 0$ for the
 three-dimensional Einstein tensor $G_{\mu\nu}$
 of generic axisymmetric spacetimes.%
 \footnote{
 For a specific spacetime, it is possible that 
 $G^{\mu\nu}\nabla_\mu \widetilde{K}_\nu = 0$ also holds while the
 spacetime is not warped product.
 We can see an example of such spacetimes 
 in Ref.~\cite{Flory:2013ssa}, which is a solution of a three dimensional
 higher derivative gravity.
 }
 Thus, $G^{\mu\nu}\widetilde{K}_\nu$ is not always a locally
 conserved current unless the axisymmetric spacetime becomes a warped
 product such as nonrotating spacetimes.

 Now, we define a new vector associated with the axial Killing vector as 
 \begin{equation}
  K^\mu \equiv - \frac{1}{2}\epsilon^{\mu\alpha\beta}\nabla_\alpha \psi_\beta .
   \label{eq:extended_Kodamavector}
 \end{equation}
 It is worth noting that this vector can be decomposed into normal
 and tangential components to $\psi^\mu$ as 
 \begin{equation}
    \begin{aligned}
   K^\mu = - \frac{1}{r}\epsilon^{\mu\nu\alpha}\psi_\alpha\nabla_\nu r 
   - \frac{\beta}{2r^2}\psi^\mu ,\quad
   \beta =\epsilon^{\lambda\alpha\beta}\psi_\lambda \nabla_\alpha
     \psi_\beta .
  \end{aligned}
 \end{equation}
 Thus, two vectors $K^\mu$ and $\widetilde{K}^\mu$ are related as 
 $K^\mu = \widetilde{K}^\mu - \beta \psi^\mu /(2r^2)$.
 If $\beta = 0$, which means that the axial Killing vector $\psi^\mu$ is
 hypersurface orthogonal and the spacetime becomes warped product, then
 both vectors are identical to just the Kodama vector.

 In what follows we will examine some properties of $K^\mu$.
 By definition, $K^\mu$ itself is 
 (i) divergence-free: $\nabla_\mu K^\mu =0$.
 We can see that (ii) $K^\mu$ is tangent to $r=\text{const.}$ surfaces,
 because 
 \begin{equation}
  \begin{aligned}
   K^\mu\nabla_\mu r &= K^\mu\nabla_\mu (\psi^\nu \psi_\nu)^{1/2}
   = \frac{1}{r} \psi^\nu K^\mu \nabla_\mu \psi_\nu 
   = \frac{1}{r} \psi^\nu K^\mu \epsilon_{\alpha\mu\nu}K^\alpha = 0 .
  \end{aligned}
 \end{equation}

 The Killing equation for the Killing vector $\psi^\mu$ gives us 
 \begin{equation}
  \nabla_\mu \nabla_\alpha \psi_\beta = - R_{\alpha\beta\mu}{}^\nu
   \psi_\nu ,
 \end{equation}
 and, in three dimensions the Riemann tensor $R_{\alpha\beta\mu\nu}$
 can be expressed as%
 \footnote{
 In three-dimensional spacetime, we can directly show 
 \begin{equation*}
  \begin{aligned}
   \epsilon_{\alpha\beta\gamma}\epsilon^{\mu\nu\lambda}G^{\gamma}{}_{\lambda}
   &= -3! g_{[\alpha}{}^\mu g_{\beta}{}^\nu g_{\gamma]}{}^\lambda
   G^{\gamma}{}_{\lambda}\\
   &= - \left(
   g_{\alpha}{}^\mu g_{\beta}{}^\nu g_{\gamma}{}^\lambda
   + g_{\beta}{}^\mu g_{\gamma}{}^\nu g_{\alpha}{}^\lambda
   + g_{\gamma}{}^\mu g_{\alpha}{}^\nu g_{\beta}{}^\lambda
   - g_{\alpha}{}^\mu g_{\gamma}{}^\nu g_{\beta}{}^\lambda
   - g_{\gamma}{}^\mu g_{\beta}{}^\nu g_{\alpha}{}^\lambda
   - g_{\beta}{}^\mu g_{\alpha}{}^\nu g_{\gamma}{}^\lambda
   \right) G^{\gamma}{}_{\lambda} \\
   &= g_\alpha{}^\mu R_\beta{}^\nu - g_\beta{}^\mu R_\alpha{}^\nu
    + R_\alpha{}^\mu g_\beta{}^\nu - R_\beta{}^\mu g_\alpha{}^\nu
   - \frac{R}{2} \left(
   g_{\alpha}{}^\mu g_{\beta}{}^\nu - g_{\beta}{}^\mu g_{\alpha}{}^\nu
   \right) \\
   &= R_{\alpha\beta}{}^{\mu\nu} .
  \end{aligned}
 \end{equation*}
 } 
 \begin{equation}
  R_{\alpha\beta\mu\nu} = 
   \epsilon_{\alpha\beta\gamma}\epsilon_{\mu\nu\lambda}G^{\gamma\lambda} .
 \end{equation}
 Using the above equations, we find 
 (iii) $G^{\mu\nu}\nabla_\mu K_\nu = 0$ as follows:  
 \begin{equation}
  \begin{aligned}
   G^{\mu\nu}\nabla_\mu K_\nu &= 
   - \frac{1}{2}\epsilon_\nu{}^{\alpha\beta}G^{\mu\nu}
   \nabla_\mu \nabla_\alpha\psi_\beta \\
   &= \frac{1}{2}\epsilon_\nu{}^{\alpha\beta}G^{\mu\nu}
   R_{\alpha\beta\mu\lambda} \psi^\lambda \\
   &= G^{\mu\nu}
   \left(\epsilon_{\nu\alpha\lambda}G_\mu{}^\alpha
   - \epsilon_{\nu\alpha\mu}R_\lambda{}^\alpha
   \right)\psi^\lambda \\
   &= 0 .
  \end{aligned}
 \end{equation}

 As a result, we confirm that $K^\mu$ can always satisfy the same conditions 
 (i) $\nabla_\mu K^\mu = 0$, (ii) $K^\mu \nabla_\mu r = 0$, and (iii)
 $G^{\mu\nu} \nabla_\mu K_\nu = 0$ 
 as
 the Kodama vector in spherically symmetric spacetimes.
 This implies that the vector 
 $K^\mu$ defined by (\ref{eq:extended_Kodamavector}) is natural extension of the Kodama vector to axisymmetric
 spacetime with angular momentum in three dimensions.

 \section{quasilocal quantities}
 \label{sec:quasi-local_quantities}
 
 In spherically symmetric cases, a locally conserved current constructed
 from the Kodama vector yields a quasilocal mass as an associated charge~\cite{Kodama:1979vn,Hayward:1994bu}. 
 In this section, we will construct quasilocal quantities related to the
 extended Kodama vector. 

 In general, if a conserved current $\mathcal{J}^\mu$ satisfying 
 $\nabla_\mu \mathcal{J}^\mu = 0$ is axisymmetric, 
 we obtain 
 \begin{equation}
  0 = 
   \pounds_\psi
   \mathcal{J}^\mu = 
   \psi^\nu \nabla_\nu \mathcal{J}^\mu
   - \mathcal{J}^\nu \nabla_\nu \psi^\mu 
   = 2\nabla_\nu ( \psi^{[\nu} \mathcal{J}^{\mu ]} ) , 
 \end{equation}
 where we have used the fact that $\psi^\mu$ and $\mathcal{J}^\mu$ are
 divergence-free.
 This implies that there exists a scalar function $\mathcal{Q}$ such
 that 
 \begin{equation}
  \nabla_\mu \mathcal{Q} = 
   2\pi \epsilon_{\mu \alpha\beta} \psi^\alpha \mathcal{J}^\beta .
 \end{equation}
 By the Stokes theorem this scalar function is written in the integral form 
 \begin{equation}
  \mathcal{Q}[C] \equiv 
  \frac{1}{2\pi}\oint_C \mathcal{Q} d\phi = \int_S \mathcal{J}^\mu
  dS_\mu ,
 \end{equation}
 where $C$ is a closed Killing orbit of the axial Killing vector and $S$ is an arbitrary spacelike
 surface whose boundary is $C$.
 It turns out that scalar function $\mathcal{Q}$ is a charge
 associated with each closed Killing orbit.
 
 Since $K^\mu$ and $G^{\mu\nu}K_\nu$ are local conserved currents as
 we mentioned in the previous section, we
 obtain scalar functions $\psi^\mu \psi_\mu$ and $K^\mu K_\mu$ 
 associated with each current, respectively.
 Indeed, we can explicitly confirm that 
 \begin{equation}
  \begin{aligned}
   \nabla_\mu (\psi^\nu \psi_\nu) &=
   2 \psi^\nu \nabla_\mu \psi_\nu \\
   &= 2 K^\alpha \epsilon_{\alpha\mu\nu}\psi^\nu ,
  \end{aligned}
  \label{eq:DPsiPsi}
 \end{equation}
 and
 \begin{equation}
  \begin{aligned}
   \nabla_\mu (K^\nu K_\nu) &=
   - \nabla^\alpha \psi^\beta \nabla_\mu \nabla_\alpha \psi_\beta \\
   &= \nabla^\alpha \psi^\beta R_{\alpha\beta\mu\nu}\psi^\nu \\
   &= \nabla^\alpha \psi^\beta 
   \epsilon_{\alpha\beta\gamma}\epsilon_{\mu\nu\lambda}G^{\gamma\lambda}
   \psi^\nu \\
   &= - 2 K_\gamma G^{\gamma\lambda} \epsilon_{\lambda\mu\nu}\psi^\nu .
  \end{aligned}
  \label{eq:DKK}
 \end{equation}
 Because any linear combination of local conserved currents is also
 locally conserved, we should adopt a charge associated with energy
 current as quasilocal mass.
 Now, we define a mass function $m$ as 
 \begin{equation}
  m \equiv \frac{1}{8G_3}(- \Lambda \psi^\nu \psi_\nu + K^\nu K_\nu) ,
   \label{eq:quasilocal_M}
 \end{equation}
 where $G_3$ and $\Lambda$ denote a three-dimensional gravitational
 constant and a cosmological constant, respectively.
 By using Eqs.~(\ref{eq:DPsiPsi}) and (\ref{eq:DKK}), we obtain 
 \begin{equation}
  \begin{aligned}
   \nabla_\mu m &= \frac{1}{4G_3} \epsilon_{\mu\nu\alpha}\psi^\alpha
   (G^{\nu\beta} + \Lambda g^{\nu\beta})K_\beta \\
   &= 2\pi \epsilon_{\mu\nu\alpha}\psi^\alpha
   T^{\nu\beta} K_\beta ,
  \end{aligned}
 \end{equation}
 where we have assumed the Einstein equation 
 $G_{\mu\nu}+\Lambda g_{\mu\nu} = 8\pi G_3 T_{\mu\nu}$ in the last line.
 Thus, $m$ is a charge associated with the conserved energy current,
 $- T^{\mu\nu} K_\nu$, in Einstein gravity. 
 
 In addition, we have an angular-momentum function as 
 \begin{equation}
  j \equiv \frac{1}{8G_3}\epsilon^{\alpha\mu\nu}\psi_\alpha \nabla_\mu
   \psi_\nu
   = - \frac{1}{4G_3}\psi_\mu K^\mu ,
   \label{eq:quasilocal_AM}
 \end{equation}
 which satisfies 
 \begin{equation}
  \begin{aligned}
   \nabla_\mu j &= - \frac{1}{4G_3}\epsilon_{\mu\nu\alpha}\psi^\alpha
   G^{\nu\beta}\psi_\beta \\
   &= - 2\pi\epsilon_{\mu\nu\alpha}\psi^\alpha
   T^{\nu\beta}\psi_\beta .
  \end{aligned}
 \end{equation}
 This also implies that $j$ is a charge associated with the conserved current,
 $T^{\mu\nu}\psi_\nu$, which is nothing but a well-known conserved current with
 respect to the axial Killing vector $\psi^\mu$.
 Note that formula~(\ref{eq:quasilocal_AM}) of angular-momentum
 function agrees with that of the Komar angular momentum. 

 The above mass and angular-momentum functions are rewritten as 
 \begin{equation}
  8G_3 m = - \Lambda r^2 - \nabla_\mu r \nabla^\mu r
   + \frac{(4 G_3 j)^2}{r^2} .
 \end{equation}
 It turns out that the mass function consists of the conventional
 Misner-Sharp mass in spherically symmetric cases and the
 angular-momentum term.%
 \footnote{
 Precisely speaking, we can add an arbitrary constant term to 
 definition of quasilocal mass (\ref{eq:quasilocal_M}).
 This constant is related to a value of mass in vacuum.
 Here, we have set $m = - 1/8G_3$ in pure AdS$_3$.
 }
 In the Ba\~{n}ados-Teitelboim-Zanelli (BTZ) solution~\cite{Banados:1992wn}, which is
 stationary, axisymmetric vacuum solution with a negative cosmological
 constant ($\Lambda < 0$), the spacetime metric is given by 
 \begin{equation}
  ds^2 = - f(r) dt^2 + \frac{dr^2}{f(r)}
   + r^2 \left(d\phi - \frac{4G_3J}{r^2}dt\right)^2 ,\quad
   f(r) = - \Lambda r^2 - 8G_3 M + \frac{16G_3^2J^2}{r^2} ,
 \end{equation}
 where $M$ and $J$ denote mass and angular-momentum parameters, respectively.
 In this case quasilocal mass $m$ and quasilocal angular momentum $j$
 defined by Eqs.~(\ref{eq:quasilocal_M}) and (\ref{eq:quasilocal_AM}) become
 constants equal to the mass and angular-momentum parameters everywhere as $m=M$ and $j=J$.
 Moreover, 
 the extended Kodama vector coincides with usual
 time-translational Killing vector,
 $K^\mu = (\partial/\partial t)^\mu$.

 \section{Conclusion}

 In this paper we have proposed extending the notion of the Kodama vector to
 axisymmetric, three-dimensional spacetimes with nonzero angular momentum.
 The extended Kodama vector 
 $K^\mu = -\frac{1}{2}\epsilon^{\mu\alpha\beta}\nabla_\alpha\psi_\beta$, 
 which is associated with an axial Killing vector $\psi^\upsilon$, satisfies the same conditions as the
 conventional Kodama vector in spherically symmetric cases.
 In particular, satisfying condition $G^{\mu\nu}\nabla_\mu K_\nu = 0$
 for the Einstein tensor makes $G^{\mu\nu}K_\nu$ divergence free for
 generic axisymmetric spacetimes.
 This provides a locally conserved energy current assuming the Einstein
 equation.
 We have shown that quasilocal mass like the Misner-Sharp mass can be
 defined as a charge associated with the locally conserved energy current.
 This quasilocal mass contains angular momentum of rotating spacetime.

 These quasilocal quantities and conservation laws were also
 pointed out in Ref.~\cite{Gundlach:2021six}, where they were useful for
 reducing the Einstein equations in numerical simulations of
 axisymmetric Einstein-perfect fluid systems.
 What we should emphasis in the present paper is that the extended
 Kodama vector guarantees the conserved
 currents and provides the quasilocal quantities as conserved charges.
 The above properties originate from purely geometrical identities for 
 three-dimensional axisymmetric spacetimes independent of field equations
 of a specific gravitational theory.
 This means that any axisymmetric, three-dimensional spacetime can admit a
 conserved current associated with the extended Kodama vector even in
 gravitational theories other than the Einstein gravity.
 The field equations will give physical meaning of the conserved current
 by connecting with energy-momentum tensor of matter in each
 gravitational theory.

 It is fascinating to generalize the extended Kodama vector to
 higher-dimensional axisymmetric spacetimes, or 
 spacetimes different from warped product, 
 other than three dimensions.
 It does not seem to be so easy, because geometrical properties in
 three-dimensional spacetimes are necessary in order to explore the
 extended Kodama vector.%
 \footnote{
 It was shown that, if a three-dimensional spacetime admits a non-null
 Killing vector field, the spacetime metric and the Ricci tensors are
 determined by two scalars characterizing the Killing vector in
 Ref.~\cite{Gurses:2010sm}.}
 However, once one can reduce to a three-dimensional axisymmetric spacetime,
 it is expected that there are a certain extended Kodama vector and locally
 conserved current associated with it.
 For example, if a cylindrical spacetime in four dimensions are
 rotating, there is expected to be a similar conserved current.

 \begin{acknowledgments}
  This work was supported by JSPS KAKENHI Grant No.~JP16K17704.
 \end{acknowledgments}

\end{document}